\documentclass[12pt]{article}

\usepackage[a4paper,margin=1in]{geometry}
\usepackage[table]{xcolor}
\usepackage{amsmath,amssymb,amsfonts}
\usepackage{graphicx}
\usepackage{booktabs}
\usepackage{enumitem}
\usepackage{caption}
\usepackage{needspace}
\usepackage{subcaption}
\usepackage{tabularx}
\usepackage{array}
\usepackage{pdflscape}
\usepackage{float}
\usepackage[none]{hyphenat}
\usepackage{multirow}
\usepackage{authblk}
\usepackage[numbers,sort&compress]{natbib}
\usepackage[hidelinks]{hyperref}

\newcolumntype{Y}{>{\raggedright\arraybackslash}X}
\newcommand{\ndfeb}{Nd$_2$Fe$_{14}$B}
\newcommand{\zrbtwo}{ZrB$_2$}
\newcommand{\Eform}{E_{\mathrm{form}}}
\newcommand{\Eseg}{E_{\mathrm{seg}}}
\newcommand{\muB}{\mu_{\mathrm B}}

\title{First-Principles Thermodynamics of Zr--B Segregation at Grain Boundaries in Recycled Nd$_2$Fe$_{14}$B}

\author[1]{Avik Mahata\thanks{Corresponding author: \href{mailto:mahataa@merrimack.edu}{mahataa@merrimack.edu}}}
\author[2]{Miha Zakotnik}
\affil[1]{Department of Mechanical and Electrical Engineering, Merrimack College, North Andover, MA, USA}
\affil[2]{Mendelan LLC, Hudson, OH, USA}
\date{}

\begin{document}

\maketitle

\begin{abstract}
Grain-boundary chemistry is central to the coercivity and thermal stability of \(\mathrm{Nd_2Fe_{14}B}\) permanent magnets, particularly in recycled magnets where recovery of the hard-magnetic phase does not by itself restore the intergranular microstructure. Our recent experimental study showed that nanoscale ZrB$_2$ precipitates can emerge at grain boundaries and triple junctions during recycling despite an overall Zr concentration of only about $0.1$~at.\%, where they are associated with boundary stabilization and suppression of grain coarsening. Here we use spin-polarized density-functional theory with a Hubbard correction (DFT$+U$) to determine the atomistic thermodynamics underlying this preferential localization. A systematic set of composition-matched bulk/grain-boundary DFT$+U$ supercells provides a common correlated-electron description across boundary chemistries. We compare B, Zr, Dy, and ZrB$_2$-like local configurations in bulk and grain-boundary environments of \(\mathrm{Nd_2Fe_{14}B}\). Excess B is strongly stabilized at the boundary, while Zr also shows an independent thermodynamic preference for the interfacial region. When Zr and B are combined in a ZrB$_2$-like configuration, the boundary preference is retained, indicating that the interface remains favorable as Zr--B coordination develops. These results establish a thermodynamic pathway for the co-localization of Zr and B prior to ZrB$_2$ formation. Magnetic-state analysis further shows that all compared structures remain within the same high-moment Fe-sublattice regime, and boundary-localized defects generally perturb the normalized magnetization less than their matrix counterparts. The calculations therefore provide a first-principles explanation for why Zr--B chemistry concentrates at intergranular regions and how such boundary-localized states can support the microstructural stability required for high-coercivity recycled Nd--Fe--B magnets.
\end{abstract}

\vspace{1em}

\section{Introduction}
\label{sec:introduction}

Nd--Fe--B magnets combine high saturation magnetization with large magnetocrystalline anisotropy and remain the dominant high-energy-product permanent magnets for traction motors, wind-power systems, robotics, and other electrification technologies \cite{sagawa1984new,gutfleisch2010magnetic,coey2012permanent}. Their coercivity is controlled not only by the \ndfeb{} grains but also by the chemistry, structure, and magnetic character of the grain-boundary (GB) and intergranular (IG) regions separating them \cite{hono2012strategy,sepehriamin2012grain,sasaki2016structure,woodcock2012understanding}. Grain-boundary diffusion and related microstructure-engineering strategies can reduce intergranular exchange coupling, suppress unfavorable grain growth, and substantially raise coercivity \cite{sepehriamin2013mechanism,tang2026high, ding2020microstructure}. The intergranular network is therefore not merely a secondary phase; it is a key microstructural feature that determines how effectively the intrinsic anisotropy of the \(\mathrm{Nd_2Fe_{14}B}\) grains is translated into macroscopic coercivity. Recent work combining microscopy, first-principles calculations, and micromagnetic modeling has further shown that the magnetic character of thin intergranular phases can be as important as local anisotropy changes in controlling reversal and coercivity \cite{tang2023unveiling}. Zr also has an established microstructural role in Nd--Fe--B processing. Itakura \textit{et al.} observed fine plate-like Zr--B precipitates derived from ZrB$_2$ on Nd$_2$Fe$_{14}$B growth surfaces and attributed suppression of abnormal grain growth to precipitate pinning together with changes in the Nd-rich liquid-phase distribution \cite{itakura2021zr}. More recent work has identified ZrB$_2$, ZrB, and Zr$_2$Fe precipitates in Zr-doped sintered magnets and linked Zr--B precipitation to local B-lean boundary chemistry \cite{fu2025zr}; in situ microscopy has additionally revealed structurally modified regions at Nd$_2$Fe$_{14}$B/ZrB$_2$ contacts \cite{guan2026amorphization}. These studies establish Zr--B precipitation as a recurring microstructural feature, but they do not determine the atomistic thermodynamic preference that selects intergranular sites.

This issue is especially important in magnet-to-magnet recycling, where recovery of the \ndfeb{} phase alone does not guarantee recovery of the microstructure that controls coercivity \cite{zakotnik2009multiple,zakotnik2015commercial,walton2015use,diehl2018towards}. In our earlier experimental study, we showed that grain-boundary engineering of recycled \(\mathrm{Nd_2Fe_{14}B}\) magnets can substantially improve coercivity and thermal stability, while producing nanoscale \(\mathrm{ZrB_2}\) precipitates at grain boundaries and triple junctions \cite{zakotnik2025sustainable}. The origin of this preferential intergranular precipitation, however, could not be established from the experimental observations alone. Transmission electron microscopy (TEM), scanning transmission electron microscopy with energy-dispersive X-ray spectroscopy (STEM-EDXS), and electron energy-loss spectroscopy (EELS) revealed well-crystallized nanoscale \zrbtwo{} precipitates inside intergranular regions and at contacts between neighboring \ndfeb{} grains. The particles were typically elongated, tens of nanometers thick and up to roughly $100$~nm or more in length, and frequently aligned parallel to neighboring grains along the $(001)$ direction. They occurred within a chemically heterogeneous intergranular network containing rare-earth-rich, Fe-rich, and Co--Cu-rich phases. Importantly, the bulk Zr concentration remained approximately $0.1$~at.\% in both the starting and recycled magnets, whereas distinct \zrbtwo{} precipitates were observed only after recycling \cite{zakotnik2025sustainable}. Thus the appearance of \zrbtwo{} is fundamentally a redistribution and phase-localization problem rather than a consequence of a large increase in the total Zr inventory.

Recent recycling studies reinforce the same broader point. Grain-boundary diffusion of recycled HDDR material and Cu modification of recycled sintered magnets show that diffusion pathways and the chemistry of rare-earth-rich or oxide-containing boundary phases can limit property recovery \cite{ikram2020limitations,pan2022cu}. NdCoGa-assisted remanufacturing has also been demonstrated at the 100-kg-per-batch scale, while Dy--Cu grain-boundary diffusion provides another route for recovering coercivity in recycled feedstocks \cite{mo2023ndcoga,dias2025dycu}. These developments sit within a wider effort to retain the value of rare-earth magnets through direct and alloy-based recycling rather than returning all material to separated elemental feedstocks \cite{ormerod2023sourcing}. The experimental work also identified the functional consequence of placing \zrbtwo{} in the intergranular network. The elongated precipitates were proposed to impede grain-boundary migration during the high-temperature recycling treatment and thereby suppress grain coarsening; together with Dy enrichment and Co--Cu-rich intergranular phases, this stabilized boundary architecture was associated with reduced intergranular magnetic coupling and the observed coercivity enhancement \cite{zakotnik2025sustainable}. What remained unresolved was the preceding atomistic step: why should Zr and B accumulate in these regions and develop \zrbtwo{}-like coordination there when Zr is already present at a similarly low bulk concentration in the unrecycled material? Answering that question requires separating the thermodynamics of boundary localization from the subsequent kinetics of precipitate growth and from the micromagnetic origin of coercivity. Despite the importance of interfaces, explicit first-principles studies of Nd--Fe--B microstructure interfaces remain comparatively limited. Prior calculations have examined Nd$_2$Fe$_{14}$B/NdO$_x$ interfaces and Cu site preference with supercells containing more than 200 atoms \cite{tatetsu2016cu,gohda2018electron}, surface and interface crystal-field effects \cite{tsuchiura2014crystalfield,toga2015trace}, Nd$_2$Fe$_{14}$B/(Fe,Co) exchange-spring interfaces \cite{umetsu2016interface}, and rare-earth partitioning between the 2--14--1 phase and Nd-rich phases \cite{liu2012partitioning}. A recent review emphasized that realistic first-principles calculations of permanent-magnet microstructure interfaces remain computationally demanding and are still largely restricted to selected clean or idealized interfaces \cite{gohda2021intergranular}. Several explicit large-cell interface studies therefore used open-core treatments of the rare-earth $4f$ states and generally focused on one interface chemistry at a time \cite{tatetsu2016cu,gohda2018electron}. To our knowledge, a composition-matched DFT$+U$ segregation study that treats B, Zr, Dy, and coupled Zr--B chemistry within the same explicit \ndfeb{} boundary framework has not previously been reported.

Here we address that missing thermodynamic step using spin-polarized DFT$+U$ calculations on explicit 136-atom \ndfeb{} bulk and GB models. We compare interstitial B, substitutional Zr and Dy, and a local ZrB$_2$-like motif in the matrix and at the boundary. Total magnetization, site-resolved moments, and spin-resolved Fe $d$ densities of states are used to verify that the paired bulk and boundary calculations remain in a common high-moment Fe-sublattice state. The resulting energetics show that B and Zr independently prefer the boundary and that Zr--B product-like local chemistry remains boundary-favored after co-localization. The calculations therefore explain the spatial selectivity that was missing from the experimental mechanism: a small overall Zr content can still produce locally concentrated \zrbtwo{} because the boundary is a thermodynamically preferred reservoir for both required species. When combined with the experimentally observed grain-growth suppression and intergranular magnetic isolation, this provides a connected picture from atomic segregation to the microstructure associated with high coercivity, while keeping clear that coercivity itself is not calculated by the present DFT model.

\section{Computational methodology}
\label{sec:methods}

\subsection{Electronic-structure calculations}
\label{sec:electronic}

All first-principles calculations were performed within spin-polarized density-functional theory (DFT) using the Vienna Ab initio Simulation Package (VASP) \cite{kresse1996efficiency,kresse1996efficient}. The interaction between the valence electrons and ionic cores was described using the projector-augmented-wave method \cite{blochl1994projector,kresse1999from}, and exchange--correlation effects were treated within the Perdew--Burke--Ernzerhof generalized-gradient approximation \cite{perdew1996generalized}. The localized rare-earth $4f$ states were described using the rotationally invariant DFT$+U$ formulation of Dudarev \textit{et al.} \cite{dudarev1998electron}, with an effective on-site interaction of $U_{\mathrm{eff}}=6$~eV applied to the Nd and Dy $4f$ states. This value is consistent with previous first-principles treatments of rare-earth-containing \(\mathrm{Nd_2Fe_{14}B}\) systems \cite{yu2022correlation}. A plane-wave kinetic-energy cutoff of 520~eV was used throughout. Brillouin-zone integrations were carried out using $\Gamma$-centered Monkhorst--Pack meshes \cite{monkhorst1976special}, with comparable reciprocal-space sampling maintained for the different supercells. All atomic coordinates were relaxed until the residual forces were below $0.02$~eV/\AA. The same exchange--correlation treatment, plane-wave cutoff, convergence criteria, and reciprocal-space sampling strategy were used for the corresponding bulk and grain-boundary configurations so that their total energies could be compared consistently. The magnetic calculations were initialized in a collinear high-moment configuration of the Fe sublattice, consistent with the established magnetic ground state of the 2--14--1 phase \cite{hirosawa1986magnetization}. The converged magnetic state of each structure was assessed from the total magnetic moment normalized by the number of Fe atoms, the site-resolved atomic magnetic moments, and the spin-resolved Fe $d$-projected density of states. These quantities were used to verify that the pristine and defect-containing bulk and grain-boundary structures remained within the same high-moment magnetic regime before their relative energies were interpreted as segregation or interfacial energetics. The core segregation dataset was deliberately constructed as five composition-matched bulk/GB pairs: pristine, B$_{\mathrm{int}}$, Zr$_{\mathrm{Nd}}$, Dy$_{\mathrm{Nd}}$, and the Zr--2B complex. It therefore contains ten fully relaxed 136-atom supercell states evaluated with the same DFT$+U$ treatment, supplemented by elemental reference phases, bulk ZrB$_2$, and the magnetic and electronic analyses described below. Keeping Nd, and Dy where present, $4f$ states in the correlated valence manifold throughout provides a uniform treatment of the rare-earth electronic structure across the entire bulk-to-boundary comparison rather than switching to an open-core approximation for the large interface cells \cite{tatetsu2016cu,gohda2018electron}.

\subsection{Bulk, grain-boundary, and defect models}

\begin{figure}[H]
    \centering
    \includegraphics[width=0.98\textwidth]{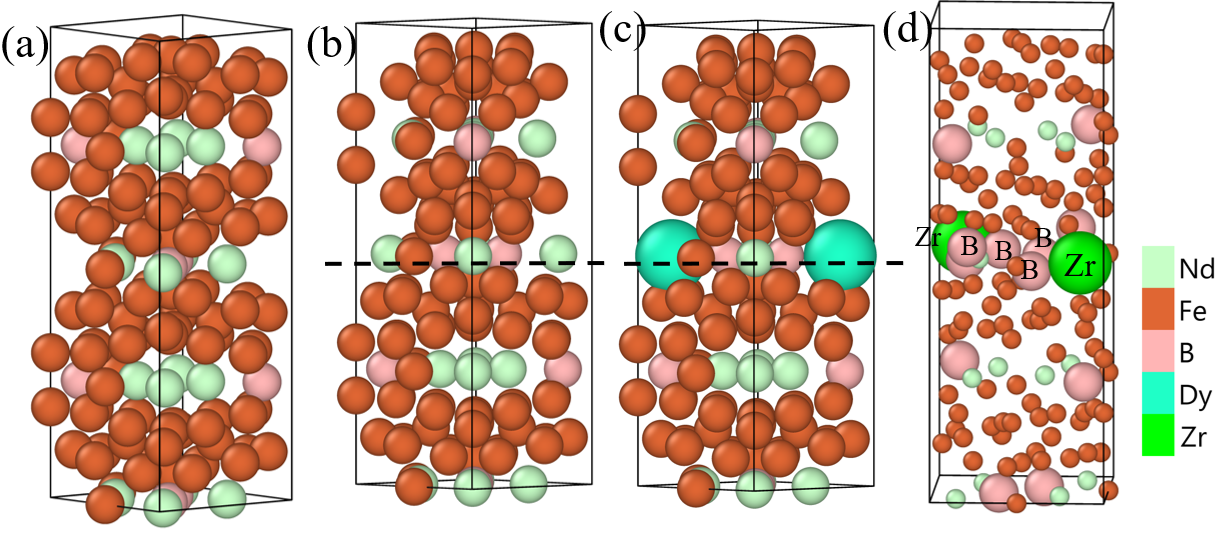}
    \caption{Atomistic models used in the DFT calculations: (a) bulk \ndfeb{}, (b) the $\Sigma1$ $(001)$ rotational interface formed by rotating the upper 68-atom slab by $90^\circ$ about $[001]$, (c) Dy at the GB, and (d) a ZrB$_2$-like Zr--B configuration at the GB. The dashed line marks the nominal interface plane.}
    \label{fig:structures}
\end{figure}

The bulk reference was a $1\times1\times2$ \ndfeb{} supercell containing Nd$_{16}$Fe$_{112}$B$_8$ (136 atoms). The GB model was constructed from the same cell by rotating the upper 68-atom half by $90^\circ$ about the $[001]$ axis, producing an idealized $\Sigma1$ $(001)$ rotational interface at the mid-plane while retaining the same composition, lattice dimensions, and atom count as the bulk reference. The model is used as a controlled boundary proxy rather than as a representation of the full distribution of high-angle grain boundaries in sintered magnets. Its $(001)$ geometry is relevant to the experimentally observed tendency of elongated \zrbtwo{} precipitates to align with $(001)$ planes \cite{zakotnik2025sustainable}; however, the present ZrB$_2$-like defect is not a crystalline precipitate and the calculations are therefore not used to infer an epitaxial orientation relationship. Periodic boundary conditions generate the corresponding companion interface at the supercell boundary. A 68-atom primitive cell was also calculated as an internal total-energy reference. Figure~\ref{fig:structures} shows the bulk reference, the idealized boundary model, and representative Dy- and Zr--B-containing boundary configurations used to define the defect comparisons.

Four principal defect classes were considered in both the bulk and grain-boundary cells. The first configuration contains an interstitial B atom, denoted B$_{\mathrm{int}}$. The second and third configurations contain Zr and Dy atoms substituting for Nd sites, denoted Zr$_{\mathrm{Nd}}$ and Dy$_{\mathrm{Nd}}$, respectively. The fourth configuration contains a local Zr--B defect complex in which Zr and B are introduced in a 1:2 stoichiometric ratio to represent ZrB$_2$-like local coordination. This configuration is used as a local precursor- or product-like motif and should not be interpreted as an explicit crystalline \zrbtwo{}/\ndfeb{} interface. Reference chemical potentials were obtained from the stable elemental phases calculated using the same electronic-structure settings. These reference states were $\alpha$-rhombohedral B, hcp Zr, hcp Dy, and dhcp Nd. Hexagonal \zrbtwo{} with the $P6/mmm$ crystal structure was also calculated as a separate reference phase \cite{li2010crystal}.

For a defect configuration, the element-referenced formation energy is
\begin{equation}
\Eform = E_{\mathrm{def}}-E_{\mathrm{host}}-\sum_i \Delta n_i\mu_i,
\label{eq:eform}
\end{equation}
where $\Delta n_i$ is the change in the number of atoms of species $i$ and $\mu_i$ is the corresponding reference chemical potential. The segregation energy is
\begin{equation}
\Eseg(X)=\left[E_{\mathrm{GB}+X}-E_{\mathrm{GB}}\right]
-\left[E_{\mathrm{bulk}+X}-E_{\mathrm{bulk}}\right],
\label{eq:eseg}
\end{equation}
where $X$ denotes the same defect unit in the two environments. Negative $\Eseg$ indicates energetic preference for the GB. Because the chemical-potential terms cancel in Eq.~\eqref{eq:eseg}, the segregation energy directly measures the relative site preference provided that the compared cells represent the same magnetic state. The magnetic checks in Sec.~\ref{sec:magnetic} establish this condition for the present dataset.

The same composition-matched bulk--GB energy pairs were also expressed as a grain-boundary excess energy,
\begin{equation}
\gamma_{\mathrm{GB}}^{X}
=
\frac{E_{\mathrm{GB}+X}-E_{\mathrm{bulk}+X}}{2A},
\label{eq:gammaGB}
\end{equation}
where $A$ is the cross-sectional area of one $(001)$ interface and the factor of two accounts for the two periodic interfaces in the bicrystal. The change relative to the pristine boundary is
\begin{equation}
\Delta\gamma_{\mathrm{GB}}^{X}
=
\gamma_{\mathrm{GB}}^{X}-\gamma_{\mathrm{GB}}^{0}.
\label{eq:deltagammaGB}
\end{equation}
Thus, $\Delta\gamma_{\mathrm{GB}}^{X}<0$ denotes a reduction in the excess energy of the modeled boundary. This quantity is not an independent thermodynamic observable from the segregation energy; rather, it recasts the same composition-matched bulk--GB energetics on an interfacial-area basis and provides a direct measure of thermodynamic boundary stabilization.

\section{Results and discussion}
\label{sec:results}

\subsection{Boundary excess energy and strong B segregation}
\label{sec:boron}

\begin{figure}[H]
\centering
\includegraphics[width=0.98\textwidth]{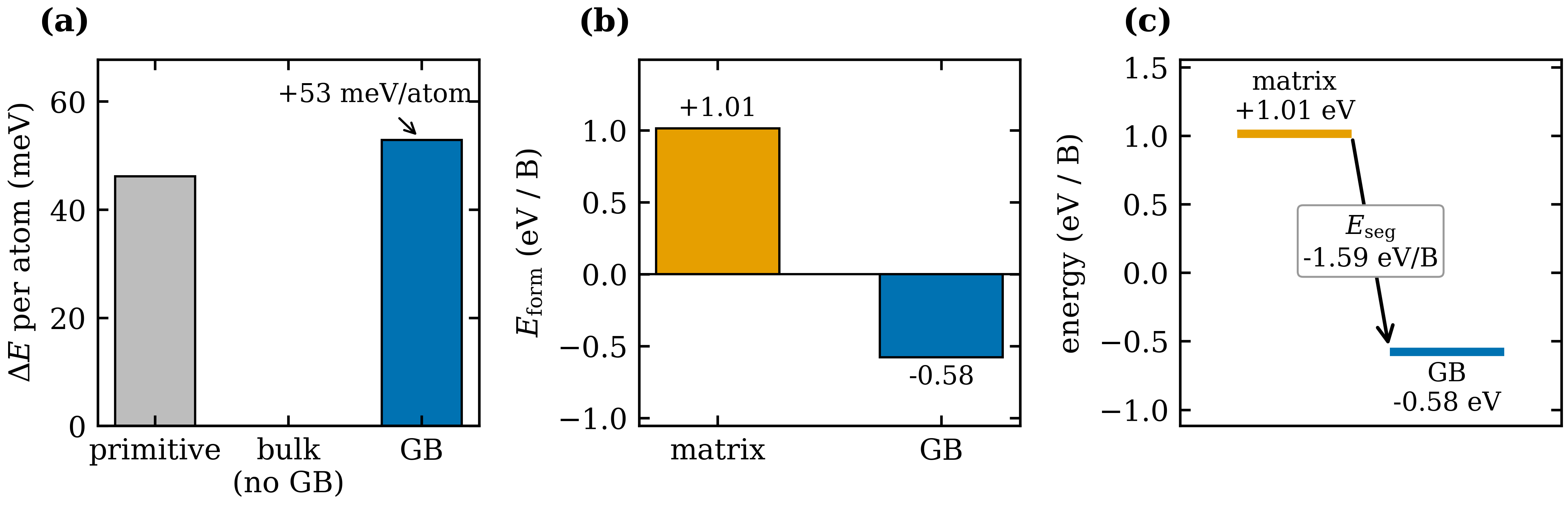}
\caption{Energetics of B segregation in Nd$_2$Fe$_{14}$B. 
(a) Energetic reference for the pristine bulk and grain-boundary environments. 
(b) Formation energy of interstitial B in the matrix and at the grain boundary. 
(c) Transfer of B from the matrix to the boundary, giving 
$\Eseg(\mathrm{B})=-1.59$~eV/B.}
\label{fig:boron}
\end{figure}

The pristine grain-boundary model has a higher total energy than the compositionally identical bulk reference, corresponding to a grain-boundary excess energy of approximately $0.75$~J\,m$^{-2}$. This value is specific to the idealized interface considered here and is not intended to represent a universal grain-boundary energy for Nd$_2$Fe$_{14}$B. Rather, it provides the reference state against which the energetic effects of boundary-localized defects are evaluated. The most pronounced segregation tendency is obtained for B. As shown in Fig.~\ref{fig:boron}, an interstitial B atom has a formation energy of $\Eform=+1.01$~eV/B in the bulk matrix, whereas the corresponding configuration at the grain boundary has $\Eform=-0.58$~eV/B. The resulting segregation energy is

\begin{equation}
\Eseg(\mathrm{B})=-1.59~\mathrm{eV/B}.
\label{eq:bseg}
\end{equation}

The negative segregation energy indicates a strong thermodynamic preference for excess B to occupy the interfacial region rather than the grain interior. In energetic terms, transferring the same B defect from the matrix to the grain boundary lowers the energy by $1.59$~eV. This preference is particularly relevant to the experimentally observed formation of ZrB$_2$, because development of a Zr--B-rich intergranular region requires not only redistribution of Zr but also a local source of B. The calculated B segregation is consistent with the EELS measurements from our earlier experimental study, which identified B within the elongated Zr-rich precipitates formed in the intergranular regions \cite{zakotnik2025sustainable}. The calculation does not establish a complete nucleation pathway for crystalline ZrB$_2$, but it shows that the grain boundary provides a strong thermodynamic sink for excess B. Such enrichment can therefore supply the local B concentration required for subsequent Zr--B coordination and precipitation without requiring an increase in the overall B content of the magnet.

\subsection{Magnetic-state consistency of the segregation calculations}
\label{sec:magnetic}

\begin{figure}[H]
\centering
\includegraphics[width=0.98\textwidth]{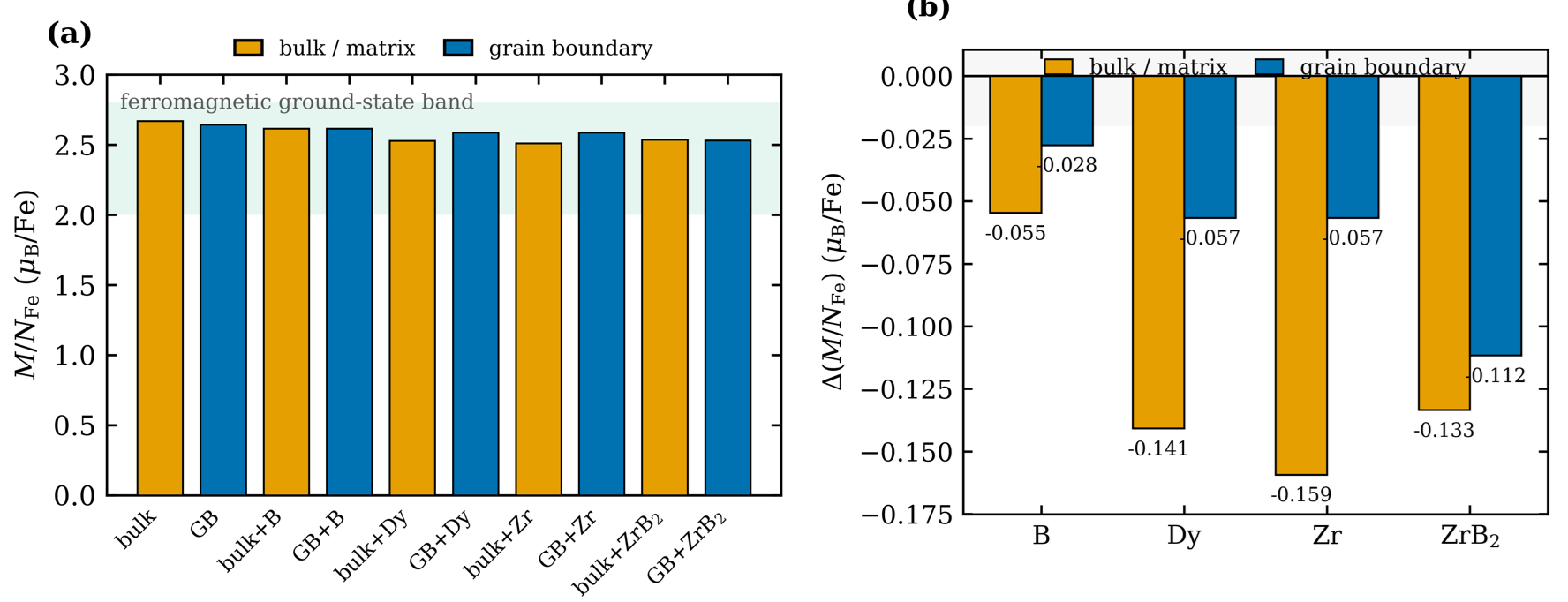}
\caption{Magnetic-state validation. (a) Total magnetization normalized by the number of Fe atoms for the pristine, B-, Dy-, Zr-, and ZrB$_2$-containing bulk and GB cells. All configurations remain on the same high-moment Fe-sublattice branch. (b) Change in normalized magnetization relative to the corresponding pristine host, $\Delta M_{\mathrm{norm}}=(M/N_{\mathrm{Fe}})_{\mathrm{defect}}-(M/N_{\mathrm{Fe}})_{\mathrm{pristine}}$. For each chemistry, the boundary-localized configuration produces a smaller magnetic perturbation than the corresponding matrix configuration. (b) is a magnetic-compatibility diagnostic and not a direct measure of coercivity.}
\label{fig:magnetic}
\end{figure}

The quantitative interpretation of segregation energies requires that bulk and GB defect cells be compared within the same electronic and magnetic regime. Figure~\ref{fig:magnetic} summarizes this validation. Panel (a) shows the total magnetic moment normalized by the number of Fe atoms. The pristine bulk and GB references and all B-, Dy-, Zr-, and ZrB$_2$-containing cells remain on the same high-moment Fe-sublattice branch; no low-moment Zr or ZrB$_2$ solutions are present in the converged dataset used for the energetics.

Panel (b) asks a different question: how strongly does each defect perturb the magnetic state relative to the corresponding pristine host environment? We define
\begin{equation}
\Delta M_{\mathrm{norm}}=
\left(M/N_{\mathrm{Fe}}\right)_{\mathrm{defect}}-
\left(M/N_{\mathrm{Fe}}\right)_{\mathrm{pristine}},
\label{eq:deltam}
\end{equation}
using the pristine bulk reference for matrix defects and the pristine GB reference for boundary defects. All four chemistries produce only modest changes compared with their respective host structures. More importantly, the GB-localized B, Dy, Zr, and ZrB$_2$-like configurations are consistently closer to their pristine reference than the corresponding matrix configurations. The contrast is strongest for Dy and Zr. This does not constitute a coercivity calculation and $\Delta M_{\mathrm{norm}}$ should not be interpreted as a local Fe moment; rather, it shows that accommodating these chemistries at the interface is magnetically less disruptive to the total high-moment state than forcing the same defects into the grain interior. This magnetic compatibility is important for interpreting the segregation energies and is consistent with a picture in which Zr--B chemistry is confined to intergranular regions while the hard-magnetic \ndfeb{} grains retain their characteristic Fe-sublattice magnetization.

\begin{figure}[H]
\centering
\includegraphics[width=0.98\textwidth]{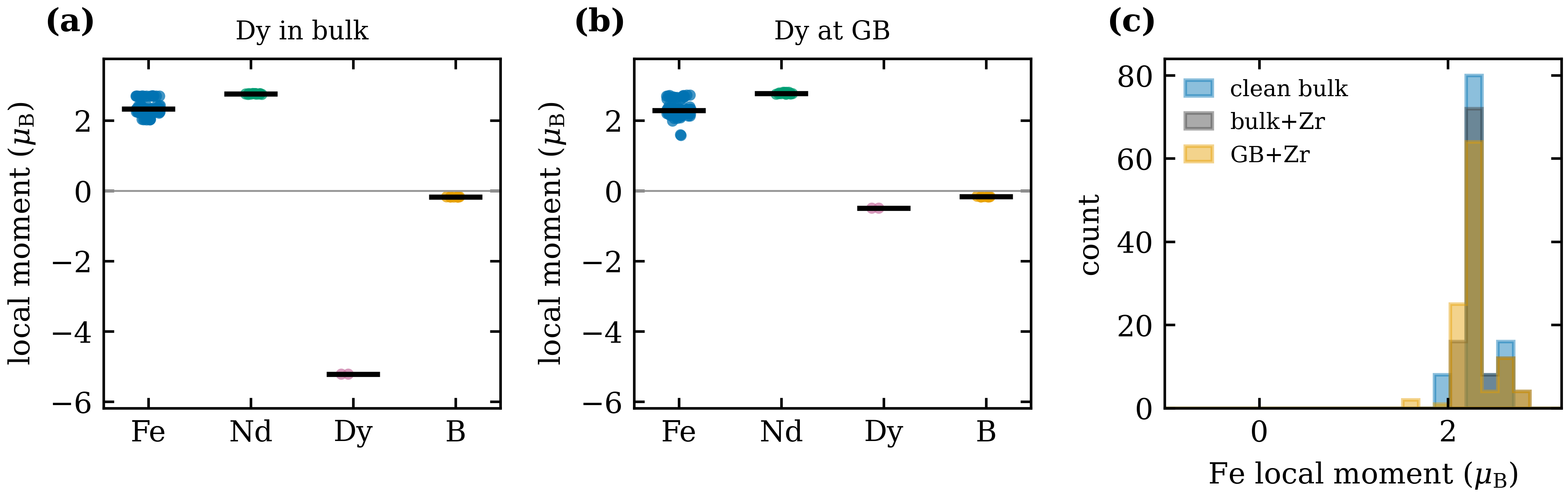}
\caption{Site-resolved magnetic response. (a) Local spin moments for Dy in the bulk cell and (b) at the GB. The Dy moment is strongly antiparallel in the bulk configuration but is reduced in magnitude at the boundary, while the Fe sublattice remains high-moment in both cases. (c) Fe local-moment distributions for pristine bulk, bulk$+$Zr, and GB$+$Zr; their strong overlap confirms magnetic consistency of the Zr segregation comparison.}
\label{fig:localmoments}
\end{figure}

The site-resolved moments provide a second, independent check, as shown in Fig.~\ref{fig:localmoments}. For Dy substitution, Fe and Nd retain predominantly positive local spin moments in both environments. The Dy local moment is strongly antiparallel in the bulk configuration, approximately $-5.2~\muB$, but is substantially reduced in magnitude at the GB. This demonstrates that the local boundary environment can modify the rare-earth magnetic response even when the total cell remains in the same high-moment regime. For Zr, the Fe-moment histograms for pristine bulk, bulk$+$Zr, and GB$+$Zr substantially overlap [Fig.~\ref{fig:localmoments}(c)], confirming that the Zr-containing configurations preserve the characteristic Fe-sublattice moment in the converged states used for energetics.

\begin{figure}[H]
\centering
\includegraphics[width=0.98\textwidth]{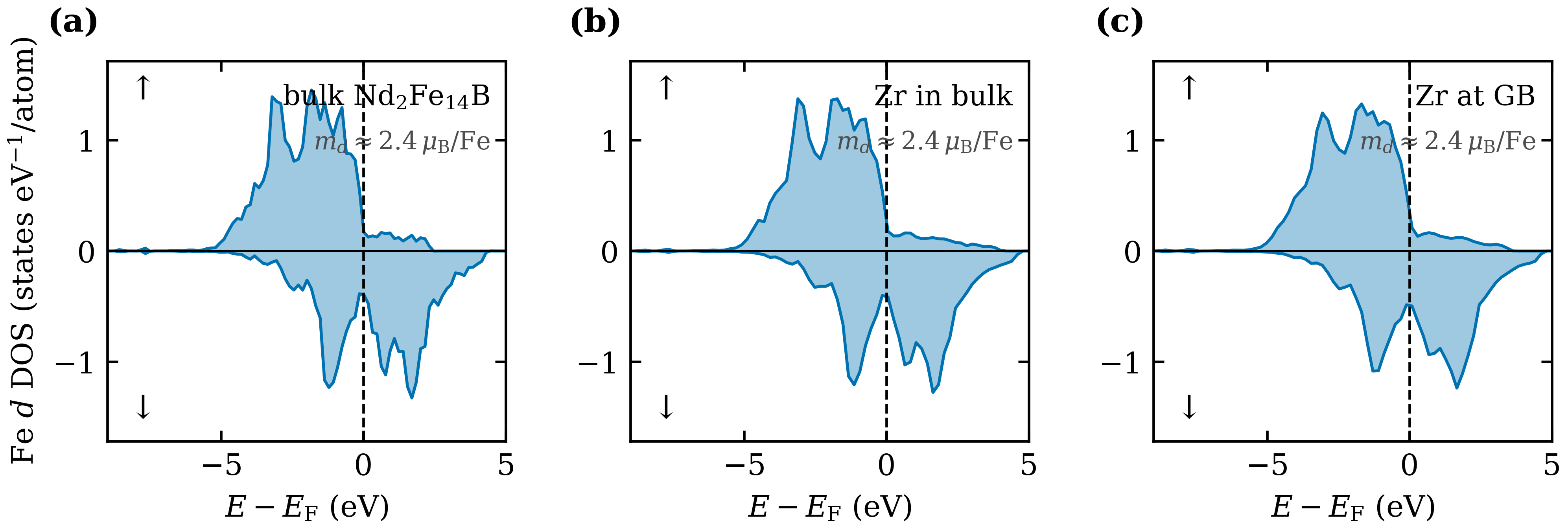}
\caption{Spin-polarized Fe $d$-projected densities of states for (a) bulk \ndfeb{}, (b) Zr in the bulk matrix, and (c) Zr at the GB. The Fermi energy is set to zero; majority and minority channels are plotted above and below the horizontal axis, respectively. The similar exchange-split line shapes and $d$-spin moments ($m_d\approx2.4~\muB$/Fe) confirm preservation of the same high-moment Fe-sublattice state.}
\label{fig:dos}
\end{figure}

Spin-resolved Fe $d$-projected DOS provides a third electronic-structure test (Fig.~\ref{fig:dos}). Bulk \ndfeb{}, Zr in bulk, and Zr at the GB exhibit closely related exchange-split Fe $d$ manifolds around the Fermi energy, with an integrated $d$-spin moment of approximately $2.4~\muB$/Fe in all three cases. The preservation of the characteristic majority/minority structure corroborates the total- and local-moment analysis. Consequently, the Zr bulk--GB energy difference can be interpreted as a genuine site-preference energy rather than an artifact of comparing inequivalent magnetic solutions.

\subsection{Segregation hierarchy of B, Zr, ZrB$_2$-like motifs, and Dy}
\label{sec:segregation}

\begin{figure}[H]
\centering
\includegraphics[width=0.98\textwidth]{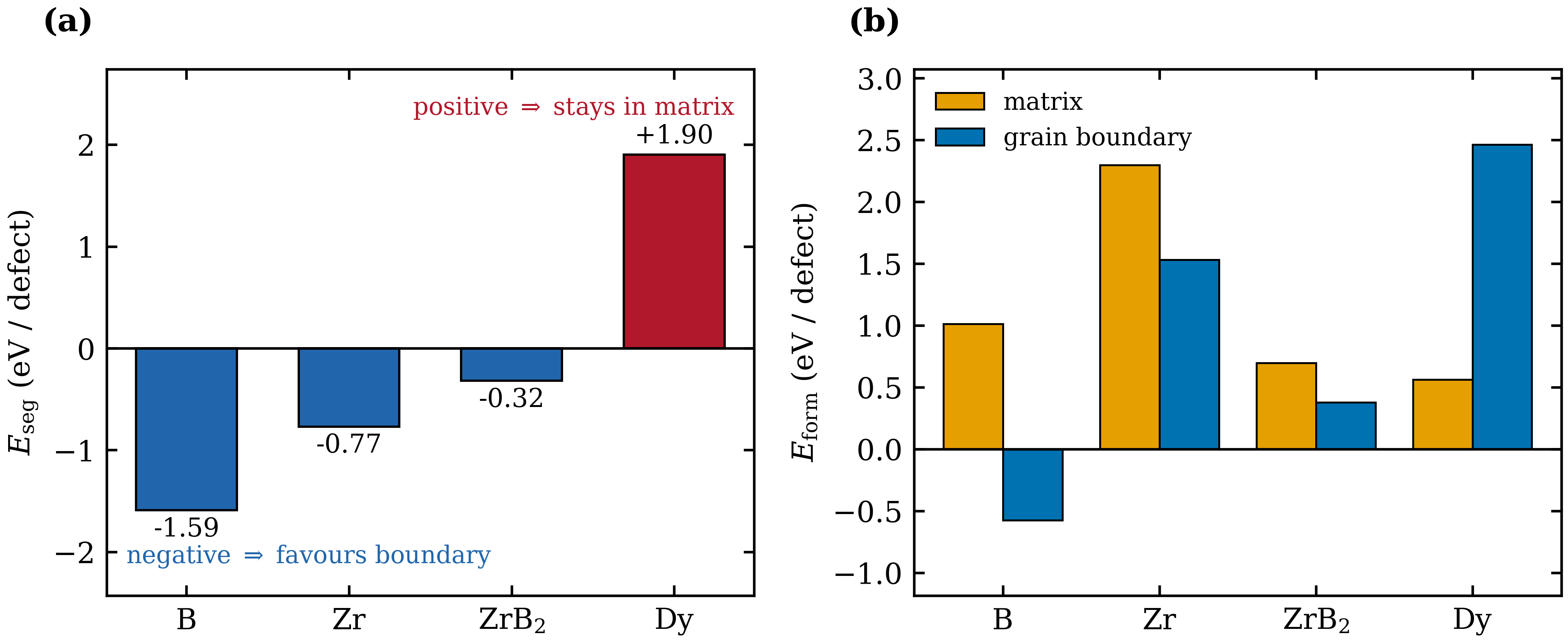}
\caption{Segregation and formation energetics. (a) Segregation energies for B, Zr, a ZrB$_2$-like defect unit, and Dy. Negative values favor the GB: $\Eseg=-1.59$, $-0.77$, and $-0.32$~eV/defect for B, Zr, and ZrB$_2$-like configurations, respectively; Dy instead gives $\Eseg=+1.90$~eV/defect. (b) Corresponding element-referenced formation energies in the matrix and at the GB. All paired configurations remain in the same high-moment magnetic regime.}
\label{fig:hierarchy}
\end{figure}

With the magnetic-state consistency established, the bulk--GB differences can be compared directly. Figure~\ref{fig:hierarchy} and Table~\ref{tab:segregation} summarize the resulting energetic hierarchy. A useful feature of this dataset is that the segregation hierarchy is not assembled from unrelated calculations: every chemistry is evaluated as a composition-matched bulk/GB pair under identical DFT$+U$ settings. The comparison therefore isolates the environmental preference of each defect while maintaining the same correlated-electron description across B, Zr, Dy, and coupled Zr--B states. Earlier first-principles Nd--Fe--B interface studies established the importance of explicit interfaces for Cu chemistry and exchange-spring coupling \cite{tatetsu2016cu,umetsu2016interface}; the present calculation extends that strategy to a multi-chemistry segregation problem. Fig.~\ref{fig:hierarchy} (a) gives the segregation energies:
\begin{align}
\Eseg(\mathrm{B}) &= -1.59~\mathrm{eV/defect}, \nonumber\\
\Eseg(\mathrm{Zr}) &= -0.77~\mathrm{eV/defect}, \nonumber\\
\Eseg(\mathrm{ZrB_2\mbox{-}like}) &= -0.32~\mathrm{eV/defect}, \nonumber\\
\Eseg(\mathrm{Dy}) &= +1.90~\mathrm{eV/defect}.
\label{eq:hierarchy}
\end{align}
For B, ``per defect'' is identical to eV/B. The negative values for B, Zr, and the ZrB$_2$-like motif mean that all three are stabilized at the modeled boundary relative to the matrix. The magnitude decreases from isolated B to Zr and then to the ZrB$_2$-like unit. Dy shows the opposite tendency and is energetically stabilized in the matrix within this idealized zero-temperature model. Figure~\ref{fig:hierarchy}(b) provides complementary formation-energy information. B is unique among the considered defects in becoming favorable relative to the elemental reference at the GB, changing from $+1.01$ to $-0.58$~eV/B. Zr and the ZrB$_2$-like motif retain positive formation energies in both environments, but their formation energies are lower at the GB, consistent with their negative segregation energies. Dy instead becomes more costly at the GB. The distinction between formation and segregation energies is important: a negative segregation energy establishes a boundary preference, but it does not by itself imply spontaneous precipitation from elemental reservoirs.

\begin{table}[H]
\caption{Calculated segregation hierarchy and magnetic-state interpretation. Negative $E_{\mathrm{seg}}$ indicates a preference for the modeled GB.}
\label{tab:segregation}
\centering
\small
\renewcommand{\arraystretch}{1.20}
\setlength{\tabcolsep}{5pt}

\begin{tabularx}{\textwidth}{@{}
>{\raggedright\arraybackslash}p{2.7cm}
>{\centering\arraybackslash}p{2.4cm}
>{\centering\arraybackslash}p{3.0cm}
>{\raggedright\arraybackslash}X
@{}}
\toprule

\textbf{Defect} &
\textbf{\begin{tabular}[c]{@{}c@{}}$E_{\mathrm{seg}}$\\(eV/defect)\end{tabular}} &
\textbf{Magnetic state} &
\textbf{Physical interpretation} \\

\midrule

$\mathrm{B_{int}}$ &
$-1.59$ &
\shortstack{High-moment in\\bulk and GB} &
Strongest segregation; excess B is stabilized at the interface \\

$\mathrm{Zr_{Nd}}$ &
$-0.77$ &
\shortstack{High-moment in\\bulk and GB} &
Direct thermodynamic preference of Zr for the interface \\

ZrB$_2$-like unit &
$-0.32$ &
\shortstack{High-moment in\\bulk and GB} &
Zr--B product-like local chemistry remains interface-favored \\

$\mathrm{Dy_{Nd}}$ &
$+1.90$ &
\shortstack{High-moment in\\bulk and GB} &
No equilibrium preference for this idealized interface \\

\bottomrule
\end{tabularx}
\end{table}

The negative Zr segregation energy establishes a direct thermodynamic driving force toward the GB: the boundary lowers the Zr defect energy relative to the matrix. The weaker but still negative value for the ZrB$_2$-like motif indicates that once Zr--B coordination develops, the boundary continues to provide an energetically preferred environment. The same result can be viewed from the complementary perspective of boundary stabilization by expressing the composition-matched bulk--GB energy differences per unit interfacial area.

\begin{table}[H]
\caption{Calculated grain-boundary excess energies for the pristine and chemically modified interfaces. $\Delta\gamma_{\mathrm{GB}}=\gamma_{\mathrm{GB}}^{X}-\gamma_{\mathrm{GB}}^{0}$ is referenced to the pristine boundary. Negative $\Delta\gamma_{\mathrm{GB}}$ indicates a reduction in the excess energy of the modeled GB.}
\label{tab:gbstabilization}
\centering
\small
\renewcommand{\arraystretch}{1.18}
\setlength{\tabcolsep}{6pt}
\begin{tabularx}{0.92\textwidth}{@{}p{3.0cm}p{3.1cm}p{3.5cm}Y@{}}
\toprule
\textbf{Boundary chemistry} &
\textbf{$\gamma_{\mathrm{GB}}$ (J m$^{-2}$)} &
\textbf{$\Delta\gamma_{\mathrm{GB}}$ (J m$^{-2}$)} &
\textbf{Interfacial interpretation} \\
\midrule
Pristine & $0.75$ & $0.00$ & Reference boundary \\
B & $0.42$ & $-0.33$ & Strong reduction in GB excess energy \\
Zr & $0.59$ & $-0.16$ & Reduced GB excess energy \\
ZrB$_2$-like & $0.55$ & $-0.20$ & Zr--B local coordination remains stabilizing at the GB \\
Dy & $1.14$ & $+0.40$ & Increased GB excess energy in the present model \\
\bottomrule
\end{tabularx}
\end{table}

Table~\ref{tab:gbstabilization} shows that the pristine boundary has an excess energy of approximately $0.75$~J\,m$^{-2}$, whereas B, Zr, and ZrB$_2$-like boundary chemistry reduce this value to approximately $0.42$, $0.59$, and $0.55$~J\,m$^{-2}$, respectively. Relative to the pristine interface, the corresponding changes are $-0.33$, $-0.16$, and $-0.20$~J\,m$^{-2}$. Dy produces the opposite response, increasing the calculated excess energy to approximately $1.14$~J\,m$^{-2}$. These values provide an interfacial interpretation of the segregation hierarchy: B and Zr are not only preferentially accommodated at the boundary, but their boundary-localized configurations reduce the energetic penalty associated with the modeled interface, and Zr--B product-like coordination retains this lower-energy interfacial character after co-localization. Because $\Delta\gamma_{\mathrm{GB}}$ is derived from the same composition-matched energy differences that enter $\Eseg$, it should be regarded as a complementary representation of the segregation thermodynamics rather than as an independent energetic mechanism. In particular, the reduced $\gamma_{\mathrm{GB}}$ values establish thermodynamic stabilization of the modeled boundary but do not by themselves constitute a grain-boundary migration barrier or a direct calculation of grain-growth kinetics.

\subsection{Thermodynamic Origin of ZrB$_2$ Boundary Localization}
\label{sec:mechanism}

In our earlier experimental study of recycled Nd--Fe--B magnets, a particularly striking microstructural feature was the appearance of well-crystallized nanoscale ZrB$_2$ precipitates within the intergranular network, even though the overall Zr concentration remained essentially unchanged at approximately $0.1$~at.\% before and after recycling \cite{zakotnik2025sustainable}. The precipitates were observed at grain boundaries and triple junctions, frequently with an elongated morphology extending along the $(001)$ direction, and were embedded within a chemically heterogeneous intergranular structure containing rare-earth-rich, Fe-rich, and Co--Cu-rich regions. Their presence was associated experimentally with a refined and stabilized boundary network, reduced grain coarsening, and improved coercivity and thermal stability. These observations established that ZrB$_2$ was not simply a consequence of introducing a larger amount of Zr during recycling. Instead, the processing route must have redistributed the small amount of Zr already present in the feedstock and brought it into local contact with B in specific intergranular regions. The experimental measurements could identify where the Zr--B-rich phase appeared and how it was incorporated into the reconstructed boundary network, but they could not determine why these locations were energetically preferred. The present DFT calculations address this unresolved point directly. Interstitial B is strongly stabilized at the modeled grain boundary relative to the grain interior, and Zr independently exhibits a negative segregation energy. The same boundary therefore provides an energetically favorable environment for both species. Moreover, when Zr and B are combined in the local 1:2 configuration used here to represent ZrB$_2$-like coordination, the boundary preference is retained. The calculations consequently provide a thermodynamic basis for the experimentally observed redistribution: the grain boundary acts as a preferential reservoir for B and Zr and remains favorable as local Zr--B coordination develops.

This interpretation also clarifies how a small global Zr inventory can produce distinct local ZrB$_2$ precipitates. During hydrogen processing, sintering, and subsequent heat treatments, Zr and B acquire sufficient mobility to redistribute through the evolving microstructure. The present calculations do not describe the diffusion kinetics or the nucleation barrier of crystalline ZrB$_2$, but they show that once these species sample different local environments, transfer from the grain interior toward the boundary is energetically favorable. The experimentally observed B signal within Zr-rich elongated particles is consistent with such a co-localization process \cite{zakotnik2025sustainable}. The grain-boundary excess-energy analysis leads to the same physical picture from an interfacial perspective. B-, Zr-, and ZrB$_2$-like configurations all reduce the excess energy of the modeled boundary relative to the pristine interface, indicating that the development of Zr--B-rich boundary chemistry is thermodynamically compatible with stabilization of the intergranular region. This result should not be interpreted as a direct grain-boundary migration barrier. The suppression of grain growth observed experimentally arises from the physical presence and morphology of the nanoscale precipitates and from the broader multiphase boundary structure, whereas the present calculations establish the energetic preference that places the relevant chemistry at those boundaries in the first place. The resulting interpretation also provides an atomistic complement to independent Zr-containing Nd--Fe--B studies. Plate-like Zr--B precipitates derived from ZrB$_2$ have been observed on Nd$_2$Fe$_{14}$B growth surfaces and shown to suppress abnormal grain growth, while more recent studies report several Zr--B-rich precipitate chemistries and structurally modified Nd$_2$Fe$_{14}$B/ZrB$_2$ interfacial regions \cite{itakura2021zr,fu2025zr,guan2026amorphization}. Those observations establish what Zr--B precipitation can do once present; the segregation energies calculated here address the preceding question of why the required Zr and B chemistry is preferentially collected at the boundary.

The magnetic calculations provide an additional connection between the atomistic energetics and the experimentally observed high-performance microstructure. All of the converged B-, Zr-, and ZrB$_2$-containing boundary configurations remain within the same high-moment Fe-sublattice regime as the pristine \ndfeb{} reference, and the boundary-localized configurations generally perturb the normalized magnetization less than the corresponding defects placed within the matrix. Thus, the thermodynamic preference for Zr--B chemistry at the interface does not require a collapse of the surrounding Fe-sublattice magnetization. This is consistent with the experimental picture in which ZrB$_2$ contributes primarily to the structural stabilization of the grain-boundary network, while Dy enrichment and Co--Cu-rich intergranular phases contribute additionally to magnetic isolation between neighboring \ndfeb{} grains \cite{zakotnik2025sustainable}. The present calculations therefore supply the missing atomistic step between processing and microstructure: thermal treatment enables redistribution, the boundary preferentially accommodates B and Zr, and local Zr--B coordination remains energetically favored as the experimentally observed ZrB$_2$-containing boundary structure develops. The calculated Dy segregation energy provides a useful contrast. Although Dy was enriched experimentally in the intergranular regions, the present idealized zero-temperature boundary gives $\Eseg(\mathrm{Dy})=+1.90$~eV. This difference emphasizes that the experimentally observed Dy distribution reflects the complete processing environment, including finite-temperature diffusion, imposed composition gradients, and interactions with rare-earth-rich and Co--Cu-rich intergranular phases, whereas B and Zr already exhibit a direct thermodynamic preference for the modeled boundary.

\section{Conclusions}
\label{sec:conclusions}

Spin-polarized DFT$+U$ calculations were used to determine the thermodynamic and magnetic consequences of B, Zr, Dy, and ZrB$_2$-like chemistry in bulk and grain-boundary environments of \(\mathrm{Nd_2Fe_{14}B}\). Methodologically, the core contribution is a common DFT$+U$ treatment of five composition-matched bulk/GB pairs, corresponding to ten 136-atom supercell states, spanning the pristine boundary and B, Zr, Dy, and coupled Zr--B chemistry while retaining the rare-earth $4f$ states as correlated valence states. To our knowledge, this combination of explicit \ndfeb{} boundary modeling, DFT$+U$, and multi-species segregation thermodynamics has not previously been reported. The calculations reveal a clear segregation hierarchy in which B exhibits the strongest preference for the modeled grain boundary, followed by Zr, while ZrB$_2$-like local coordination remains energetically favored at the interface after Zr and B are brought together. These results identify a thermodynamic pathway for the co-localization of Zr and B within intergranular regions and provide an atomistic explanation for the preferential development of ZrB$_2$-containing chemistry at grain boundaries and triple junctions. Expressed in terms of interfacial excess energy, B-, Zr-, and ZrB$_2$-like configurations also reduce the energy of the modeled grain boundary relative to the pristine interface, demonstrating that the same segregation process contributes to thermodynamic grain-boundary stabilization. Dy shows the opposite behavior within the present boundary model, emphasizing that different alloying species can interact with the intergranular environment through fundamentally different thermodynamic mechanisms. The energetic trends are supported by a consistent magnetic picture. All bulk and grain-boundary configurations remain within the same high-moment Fe-sublattice regime, and the site-resolved magnetic moments and spin-resolved Fe $d$ states confirm preservation of the characteristic exchange-split electronic structure. Boundary-localized defects generally perturb the normalized magnetization less than the corresponding defects accommodated within the matrix, indicating that segregation of Zr--B chemistry to the interface is compatible with retention of the high-moment magnetic state of the surrounding \(\mathrm{Nd_2Fe_{14}B}\) phase. The combined results therefore establish a first-principles mechanism in which grain-boundary segregation concentrates Zr and B within intergranular regions, stabilizes ZrB$_2$-like local chemistry, and lowers the energetic cost of the interface without destabilizing the Fe-sublattice magnetism. This thermodynamic and magnetic compatibility provides a microscopic basis for understanding how Zr--B-rich boundary structures can contribute to the microstructural stability required for high-coercivity \(\mathrm{Nd_2Fe_{14}B}\) magnets.

\section*{Data availability}
The relaxed structures and the correlated output files are available from the authors on reasonable request.

\section*{Acknowledgements}
This work was supported by the Department of Mechanical and Electrical Engineering at Merrimack College. The authors acknowledge the use of computational resources at the Massachusetts Green High Performance Computing Center (MGHPCC). This research also benefited from high performance computing allocations provided by the National Science Foundation through ACCESS (awards MAT250103). Additional computational resources were supported by Argonne National Laboratory under the Director's Discretionary allocation for the project GNNMD. Further support was provided through a National Science Foundation MRI Award to Wilkes University (Award No.\ 1920129), which contributed essential computational infrastructure for this study.

\bibliographystyle{unsrtnat}
\bibliography{references}

\end{document}